\documentclass[showpacs,preprintnumbers,amsmath,amssymb,floatfix]{revtex4-1}

\usepackage{graphicx}

\usepackage{dcolumn}

\usepackage{bm}

\begin{document}

\title{Particle motion around   charged black holes
in generalized dilaton-axion gravity}

\author{  Susmita Sarkar }
\email{  susmita.mathju@gmail.com } \affiliation{Department of
Mathematics, Jadavpur University, Kolkata 700 032, West Bengal,
India}

\author{Farook Rahaman}
\email{rahaman@associates.iucaa.in } \affiliation{Department of
Mathematics, Jadavpur University, Kolkata 700 032, West Bengal,
India}

\author{  Irina Radinschi }
\email{  radinschi@yahoo.com} \affiliation{Department of Physics, Gheorghe Asachi Technical University, 700050 Iasi, Romania}

\author{ Theophanes Grammenos }
\email{ thgramme@uth.gr } \affiliation{Department of Civil Engineering,
University of Thessaly, 383 34 Volos, Greece}

 \author{Joydeep Chakraborty }
\email{ joydeep_sweety2010@rediffmail.com } \affiliation{Department of Mathematics,
 Nagar College, P.O. Nagar, Dist. Mursidabad, West Bengal, India  }

\date{\today}

\begin{abstract}
The behaviour of massive and massless test particles around asymptotically flat and spherically symmetric, charged black holes in the context of generalized dilaton-axion gravity in four dimensions is studied. All the possible motions are investigated by calculating and plotting the corresponding effective potential for the massless and massive particles as well. Further,  the motion of massive (charged or uncharged) test particles in the gravitational field of  charged black holes in generalized dilaton-axion gravity for the cases of static and non-static equilibrium  is investigated by applying the Hamilton-Jacobi approach.

\end{abstract}

\maketitle

\section{Introduction}
Recently,  scientists have focused their attention to the black hole solutions in various alternative theories of gravity, particularly theories of gravitation with background scalar and pseudo-scalar fields.  In the low energy effective action, usually string theory based-models are comprised of two massless scalar fields, the dilaton and the axion (see, e.g., [1]). Sur, Das and SenGupta [2] employed the dilaton and axion fields coupled to the electromagnetic field in a more generalized coupling with Einstein and Maxwell theory in four dimensions in the low energy action. Exploiting this new idea, they have found asymptotically flat and non-flat dilaton-axion black hole solutions. The vacuum expectation values of the various moduli of compactification are responsible for these couplings. These black hole solutions have   been studied extensively in the literature,  e.g., their thermodynamics has been investigated [3],  	
thin-shell wormholes have been constructed  from charged black holes in generalized dilaton-axion gravity [4],
the energy   of charged black holes in generalized dilaton-axion gravity has been calculated [5],  the statistical entropy of a charged dilaton-axion black hole has been examined [6], and the  superradiant instability of a dilaton-axion black hole under scalar perturbation has been investigated [7].
 Among various properties of such black hole solutions, a subject of great interest is the study of the behaviour of a test particle in the gravitational field  of such black holes.

In this paper, we study the behaviour of the time-like and null geodesics  in the gravitational field of a charged black hole in generalized dilaton-axion gravity. The solution under study describes an asymptotically flat black hole and the motions of both massless and massive particles are analyzed. The effective potentials are calculated and plotted for various parameters in the cases of circular and radial geodesics. The motion of a charged test particle in the gravitational field of a charged black hole in generalized dilaton-axion gravity is also investigated using the Hamilton-Jacobi  approach.

The present paper has the following structure: in Section 2 the charged black hole metric in generalized dilaton-axion gravity is presented. Section 3 focuses on the geodesic equation in the cases of massless particle motion ($L=0$) and massive particle motion ($L=-1$). In Section 4 the effective potential is studied in both cases of the massless and the massive particle. Section 5 is devoted to the study of the motion of a test particle in static equilibrium as well as in non-static equilibrium. For the latter case, a chargeless ($e=0$) and a charged test particle are considered. Finally, in  Section 6, the results obtained in this paper are discussed.

\section{Charged black hole metric in generalized dilation-axion gravity}

Recently, Sur, Das and SenGupta [2] have discovered a new black hole solution for the Einstein-Maxwell scalar field system inspired by low energy string theory. In fact, they have considered a generalized action in which
two scalar fields are minimally coupled to the Einstein-Hilbert-Maxwell field in four dimensions (in the Einstein frame, see, e.g. [8] and [9]) having the form
\begin{equation}
I=\frac{1}{2\kappa }\int d^{4}x\sqrt{-g}\left[ R-\frac{1}{2}\partial _{\mu
}\varphi \partial ^{\mu }\varphi -\frac{\omega (\varphi )}{2}\partial _{\mu
}\zeta \partial ^{\mu }\zeta -\alpha (\varphi ,\zeta )F_{\mu \nu }F^{\mu \nu }-\beta (\varphi
,\zeta )F_{\mu \nu }\ast F^{\mu \nu }\right] ,
\end{equation}
where $\kappa =8\pi G$, R is the curvature scalar, $F_{\mu \nu }$ describes the
Maxwell field strength and $\varphi$, $\zeta$ are two massless scalar / pseudo-scalar
fields depending only on the radial coordinate $r$ which are coupled to the Maxwell field through the functions $\alpha $ and $\beta $. Here, $\zeta$ acquires a non minimal kinetic term of the form $\omega (\varphi )$ due to its interaction with $\varphi$ ($\varphi$, $\zeta$ can be identified with the scalar dilaton field and the pseudo-scalar axion field respectively), while $\ast F^{\mu \nu }=\frac{1}{2}\varepsilon^{\mu\nu\kappa\lambda}F_{\kappa\lambda}$ is the Hodge-dual Maxwell field strength.

Indeed, with the action described by Eq. (1), a much wider class of black hole solutions has been found, whereby two types of metrics, asymptotically flat and
asymptotically non flat, for the black hole solutions have been obtained.

For our study we use the asymptotically flat solution to analyze the behaviour of massive and massless test particles around a spherically symmetric, charged black hole in generalized dilaton-axion gravity. The asymptotically flat metric considered is given by
\begin{equation}
ds^{2}=-f(r)dt^{2}+\frac{dr^{2}}{f(r)}+h(r)d\Omega^{2},
\end{equation}
with
\begin{equation}
f(r)=\frac{(r-r_{-})(r-r_{+})}{(r-r_{0})^{(2-2n)}(r+r_{0})^{2n}}
\end{equation}%
and
\begin{equation}
h(r)=\frac{(r+r_{0})^{2n}}{(r-r_{0})^{(2n-2)}}.
\end{equation}

In Eqs. (3) and (4), according to [2], in order to have non-trivial $\varphi$ and $\zeta$ fields, the exponent $n$ is a dimensionless constant strictly greater than $0$ and strictly less than $1$. The other various parameters are given  as follows:
\begin{equation}
r_{\pm }=m_{0}\pm \sqrt{m_{0}^{2}+r_{0}^{2}-\frac{1}{8}\left( \frac{K_{1}}{n}%
+\frac{K_{2}}{1-n}\right) } ,
\end{equation}%
\begin{equation}
r_{0}=\frac{1}{16m_{0}}\left( \frac{K_{1}}{n}-\frac{K_{2}}{1-n}\right),
\end{equation}%
\begin{equation}
m_{0}=m-(2n-1)r_{0},
\end{equation}%
\begin{equation}
K_{1}=4n[4r_{0}^{2}+2r_{0}(r_{+}+r_{-})+r_{+}r_{-}],
\end{equation}%
\begin{equation}
K_{2}=4(1-n)r_{+}r_{-},
\end{equation}%
and
\begin{equation}
m=\frac{1}{16r_{0}}\left( \frac{K_{1}}{n}-\frac{K_{2}}{1-n}\right)
+(2n-1)r_{0},
\end{equation}%
where $m$ is the mass of the black hole and $0<n<1$. The parameters $r_{+}$ and $r_{-}$
determine the inner and outer event horizons respectively. Also, for $r=r_{0}$ there is a
curvature singularity and the parameters obey the condition $r_{0}<r_{-}<r_{+}$.

\section{Geodesic equation}

The geodesic equation for the metric (2) describing the motion in the plane $\theta =\frac{\pi}{2}$ is as follows [10]
\begin{eqnarray}
\left(\frac{dr}{d\tau}\right)^2&=&Lf(r)+E^2-\frac{J^2f(r)}{h(r)},
\\
\frac{d\phi}{d\tau}&=&\frac{J}{h(r)},
\\
\frac{dt}{d\tau}&=&\frac{E}{f(r)},
\end{eqnarray}
where $L$  is known as the Lagrangian having the values 0 for a massless particle and $-1$  for a massive particle and $E$, $J$ are constants identified as the energy per unit mass and the angular momentum respectively.

Now we proceed to discuss the motion of the massless and the massive particle for the radial geodesic.

The  radial geodesic equation ($J=0$) is
\begin{equation}
\left(\frac{dr}{d\tau}\right)^2=E^2+Lf(r).
\end{equation}
Using Eq. (13), Eq. (14) becomes
\begin{equation}
\left(\frac{dr}{dt}\right)^2=\left(f(r)\right)^2\left(1+f(r)\frac{L}{E^2}\right).
\end{equation}
Then, by inserting $f(r)$ from Eq. (3), Eq. (15) reads
\begin{equation}
\left(\frac{dr}{dt}\right)^2=\left(\frac{(r-r_-)(r-r_+)}{(r-r_0)^{(2-2n)}(r+r_0)^{2n}}\right)^2\left(1+\frac{(r-r_-)
(r-r_+)}{(r-r_0)^{(2-2n)}(r+r_0)^{2n}}\frac{L}{E^2}\right)\label{rggg}.
\end{equation}

\subsection{ Massless particle motion ($L=0$)}
For the motion of a massless particle the Lagrangian $L$  vanishes. In this case the equation for the radial geodesic (16) becomes
\begin{eqnarray}
\left(\frac{dr}{dt}\right)^2&=&\left(\frac{(r-r_-)(r-r_+)}{(r-r_0)^{(2-2n)}(r+r_0)^{2n}}\right)^2.
\end{eqnarray}
After integrating we get
\begin{eqnarray}
\pm t&=&r+\frac{ln(r-r_-)r^2}{r_--r_+}-\frac{ln(r-r_-)r_0^2}{r_--r_+}-\frac{ln(r-r_+)r_+^2}{r_--r_+}+\frac{ln(r-r_+)r_0^2}{r_--r_+}.
\end{eqnarray}
\begin{figure*}[thbp]
\begin{center}
\begin{tabular}{rl}
\includegraphics[width=7.cm]{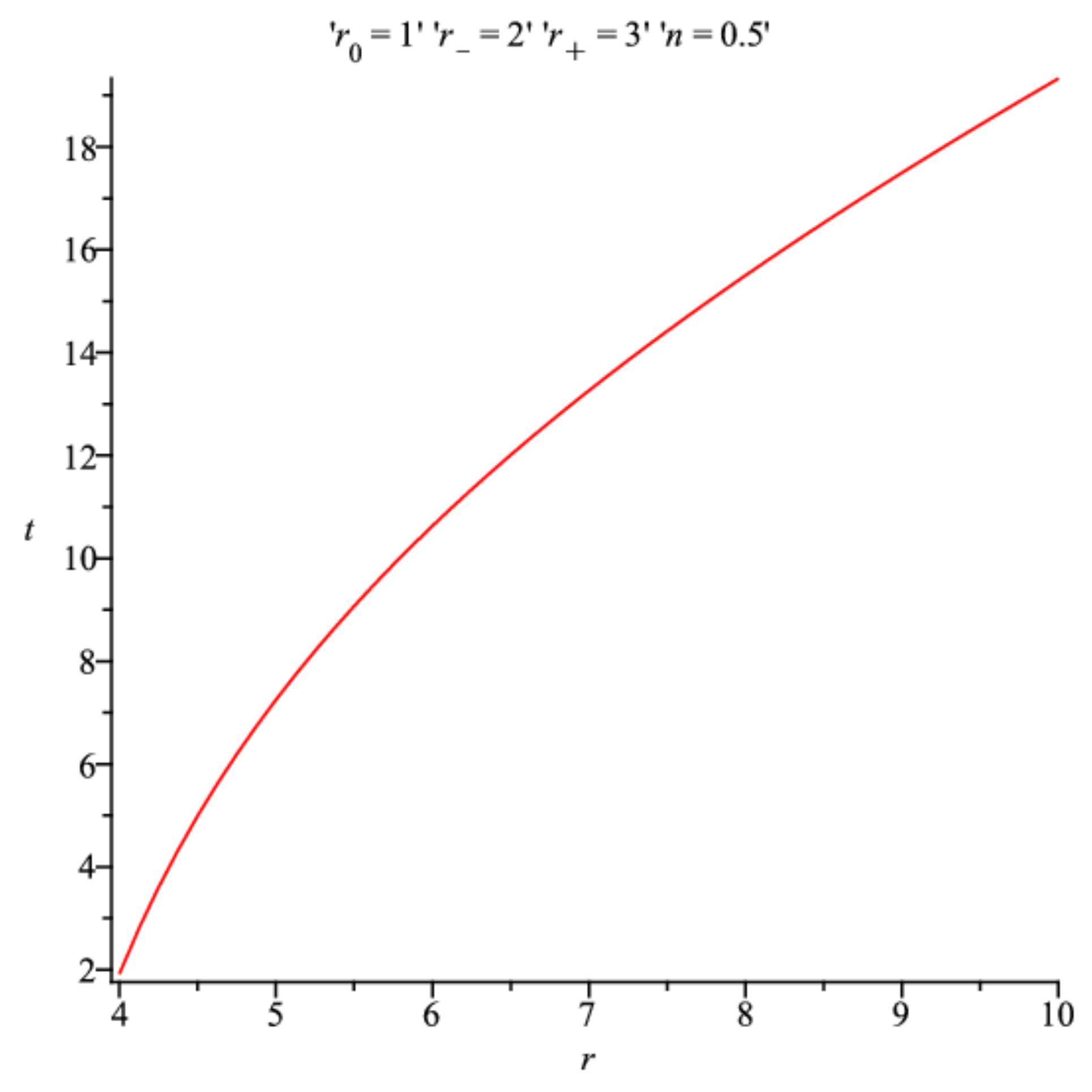}&
\includegraphics[width=7.cm]{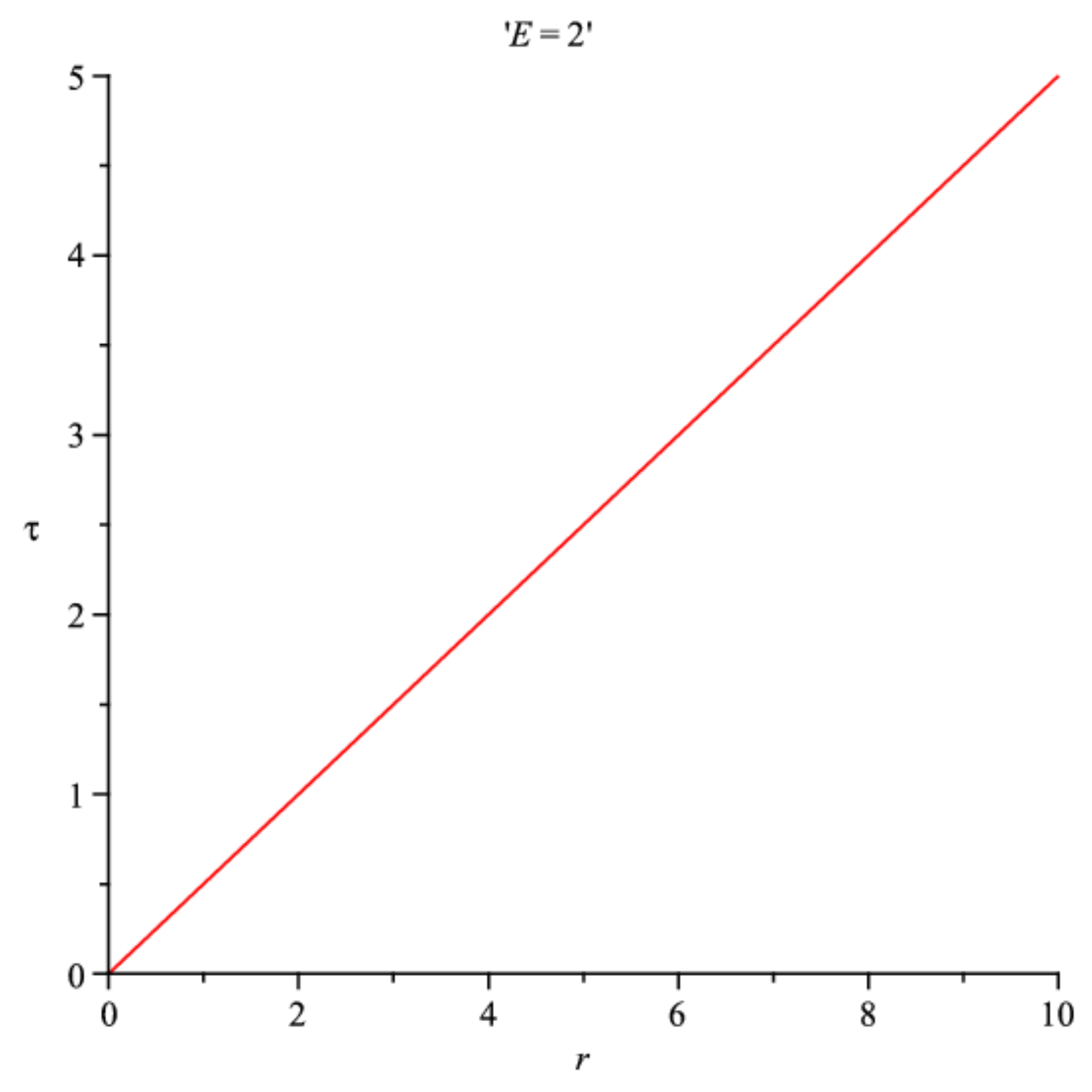}

\\
\end{tabular}
\end{center}
Fig.1: Graphs of $t-r$ (left) $\tau-r$ (right) for a massless particle.
\end{figure*}
Again from Eq. (14) we obtain for $L=0$
\begin{equation}
\left(\frac{dr}{d\tau}\right)^2=E^2\label{rt},
\end{equation}
from which we have a $\tau-r$ relationship
\begin{equation}
\pm\tau=\frac{r}{E}.
\end{equation}
In Fig. 1 (left) $t$ is plotted with respect to the radial coordinate $r$ and in Fig. 1 (right) the  proper time ($\tau$) is plotted with respect to the radial coordinate $r$ for a massless particle.

\subsection{ Massive particle motion ($L=-1$)}

For a massive particle the Lagrangian $L$  is $-1$ and from Eqs. (14) and (15) we obtain for the motion of a massive particle  the relationships between $t$ and $r$ and $\tau$ and $r$ respectively as


\begin{eqnarray}
\pm t&=&\int\frac{Edr}{\left(\frac{(r-r_-)(r-r_+)}{(r-r_0)^{(2-2n)}(r+r_0)^{2n}}\right)\left(E^2-\frac{(r-r_-)(r-r_+)}{(r-r_0)^{(2-2n)}(r+r_0)^{2n}}\right)^\frac{1}{2}}
\\
\pm \tau&=&\int\frac{(r-r_0)^{(1-n)}(r+r_0)^{n}dr}{\left(E^2(r-r_0)^{(2-2n)}(r+r_0)^{2n}-(r-r_-)(r-r_+)\right)^\frac{1}{2}}.
\end{eqnarray}


\begin{figure*}[thbp]
\begin{center}
\begin{tabular}{rl}
\includegraphics[width=7.cm]{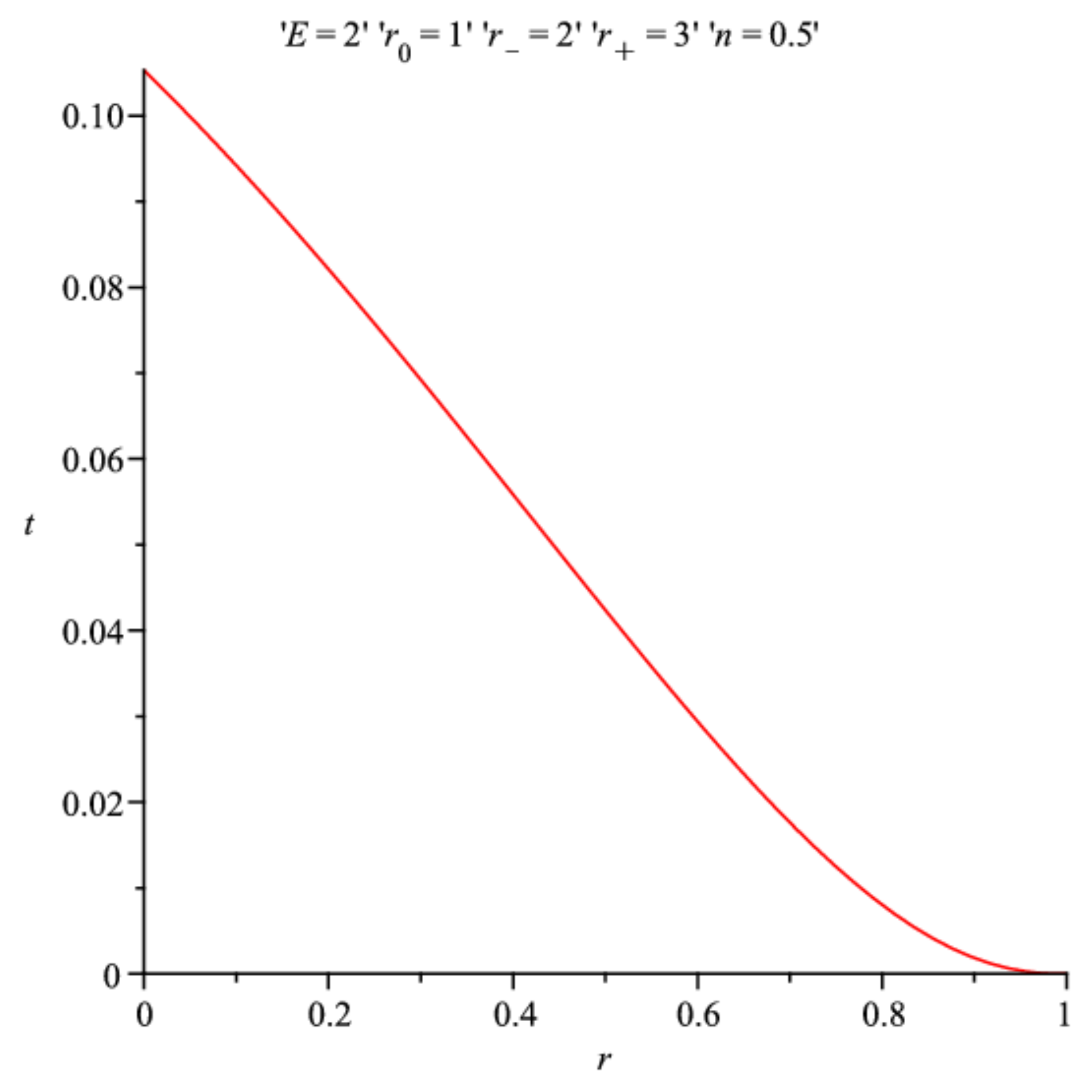}&
\includegraphics[width=7.cm]{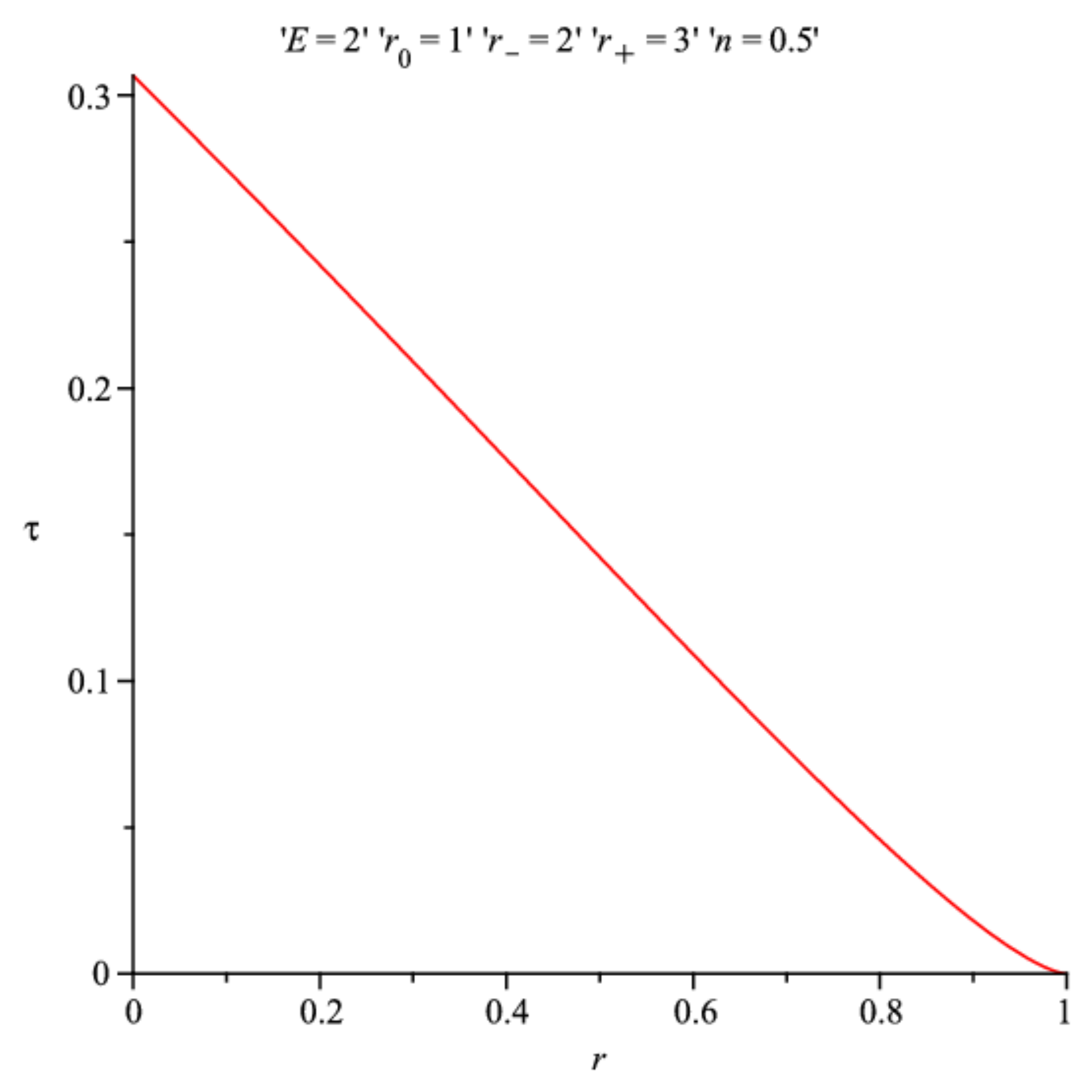}

\\
\end{tabular}
\end{center}
Fig.2: Graphs of $t-r$ (left) $\tau-r$ (right) for a massive particle.
\end{figure*}

In Fig. 2 the graphs of $t$ with respect to the radial coordinate $r$ (left) and of the proper time $\tau$ with respect to the radial coordinate $r$ (right) for a massive particle are presented.

\section{The Effective Potential}

From the geodesic Eq. (11) we have

\begin{equation}
\frac{1}{2}\left(\frac{dr}{d\tau}\right)^2=\frac{1}{2}\left[E^2-f(r)\left(\frac{J^2}{h(r)}-L\right)\right].
\end{equation}
After comparing the above equation with the well known equation $\frac{1}{2}\left(\frac{dr}{d\tau}\right)^2$+$V_{\text{\scriptsize{eff}}}$=0, we obtain the following expression for the effective potential:
\begin{equation}
V_{\text{\scriptsize{eff}}}=-\frac{1}{2}\left[E^2-f(r)\left(\frac{J^2}{h(r)}-L\right)\right]\label{ve}.
\end{equation}
From Eq. (24) one can see that the effective potential depends on the energy per unit mass, $E$ and the angular momentum, $J$.

\subsection{Massless particle case ($L=0$)}

For the radial geodesics ($J=0$), Eq. (24) yields
\begin{equation}
V_{\text{\scriptsize{eff}}}=-\frac{E^2}{2}
\end{equation}
and the particle behaves like a free particle if $E=0$.

Now we consider the circular geodesics ($J\neq$0). The corresponding effective potential is given by
\begin{equation}
V_{\text{\scriptsize{eff}}}=-\frac{E^2}{2}+\frac{J^2}{2}\frac{(r-r_-)(r-r_+)}{(r+r_0)^{4n}}.
\end{equation}

\begin{figure*}[thbp]
\begin{center}
\begin{tabular}{rl}
\includegraphics[width=8.cm]{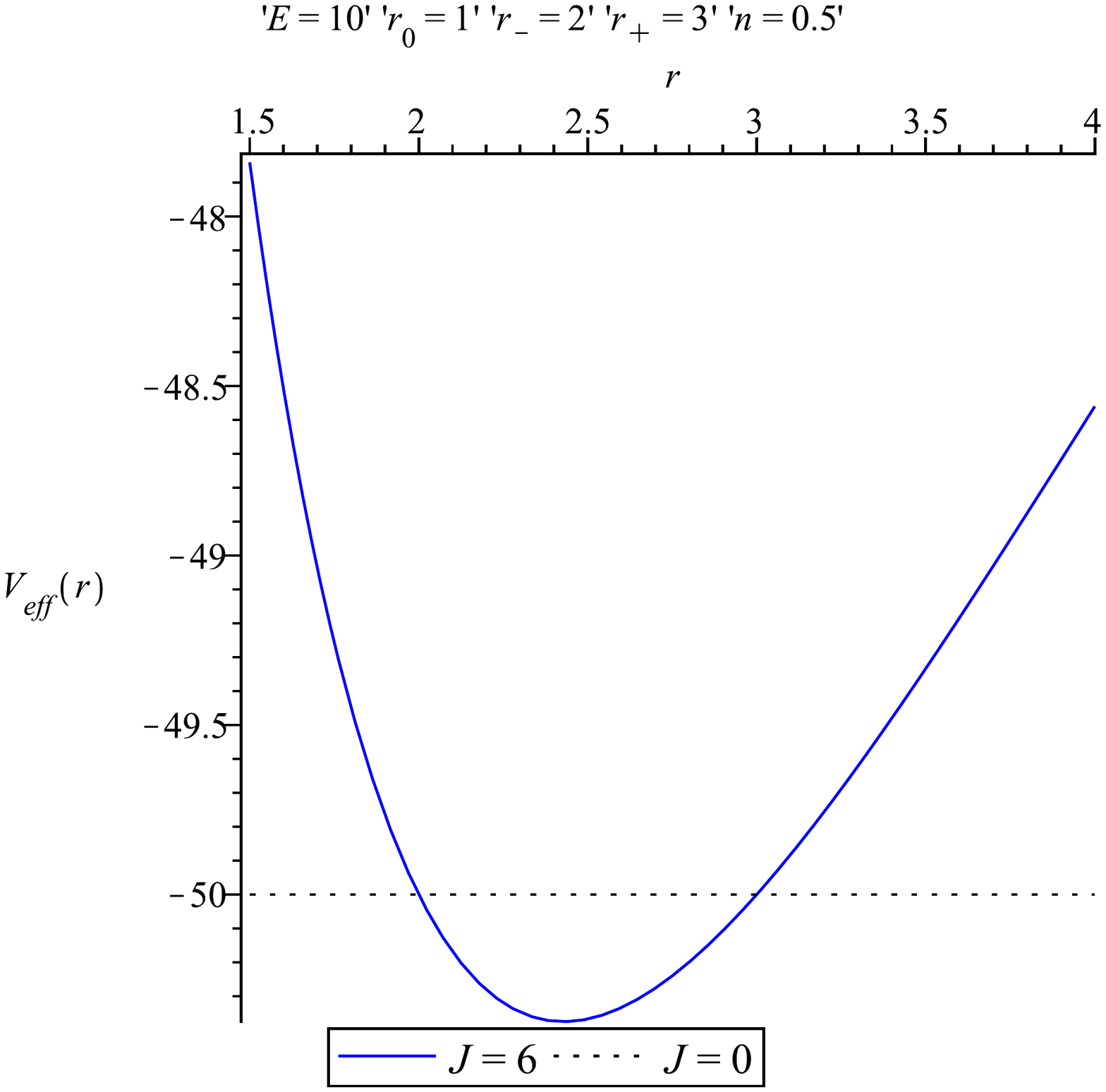}&
\includegraphics[width=7.cm]{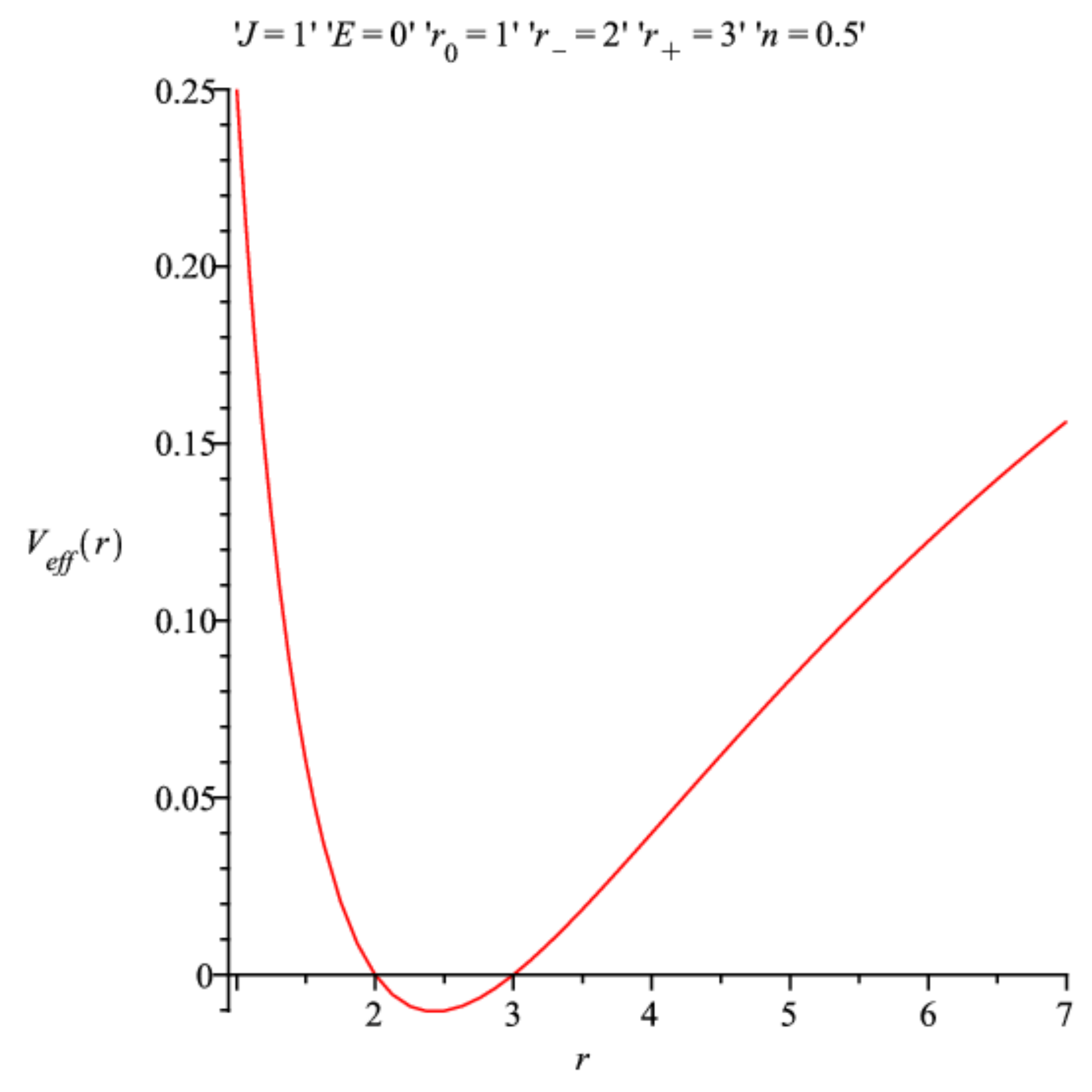}
\\
\end{tabular}
\end{center}
Fig.3: Graphs of $V_{\text{\scriptsize{eff}}}-r$ with $E= 10$, $J=6$ and $J=0$ (left) and $V_{\text{\scriptsize{eff}}}-r$ with $E= 0$ and $J=1$ (right), for a massless particle.
\end{figure*}

For $E\neq 0$ and $J=0$ we infer from Eq. (26) that the effective potential does not depend on the charge and the mass of the black hole in generalized dilaton-axion gravity. The shape of the effective potential for $ E \neq 0$ is shown in the Fig.  3 (left)  for $J=6$ (solid curve) and $J=0$ (dotted line). We notice that for a non zero value of $J$, the effective potential aquires a  minimum, implying that stable circular orbits might exist. For $J=0$, there are no stable circular orbits.

Now if we consider $E=0$ and circular geodesics, i.e. $J\neq$0, then from Eq. (26),  it is clear that the roots of the effective potential are the same as the horizon values. Further,  the effective potential is negative between its two roots, i.e. between the horizons. Hence, since the effective potential has a minimum value stable circular orbits must  exist, a conclusion that is confirmed in Fig. 3 (right).

\subsection{Massive particle case ($L=-1$)}

The effective potential for the massive particle is obtained  from Eq. (24) as
\begin{equation}
V_{\text{\scriptsize{eff}}}=-\frac{1}{2}\left[E^2-f(r)\left(\frac{J^2}{h(r)}+1\right)\right].
\end{equation}

Now, for the radial geodesics with $J=0$, $E=0$, the above equation yields
\begin{equation}
V_{\text{\scriptsize{eff}}}=\frac{1}{2}\frac{(r-r_-)(r-r_+)}{(r-r_0)^{(2-2n)}(r+r_0)^{2n}}.
\end{equation}

\begin{figure*}[thbp]
\begin{center}
\begin{tabular}{rl}
\includegraphics[width=7.cm]{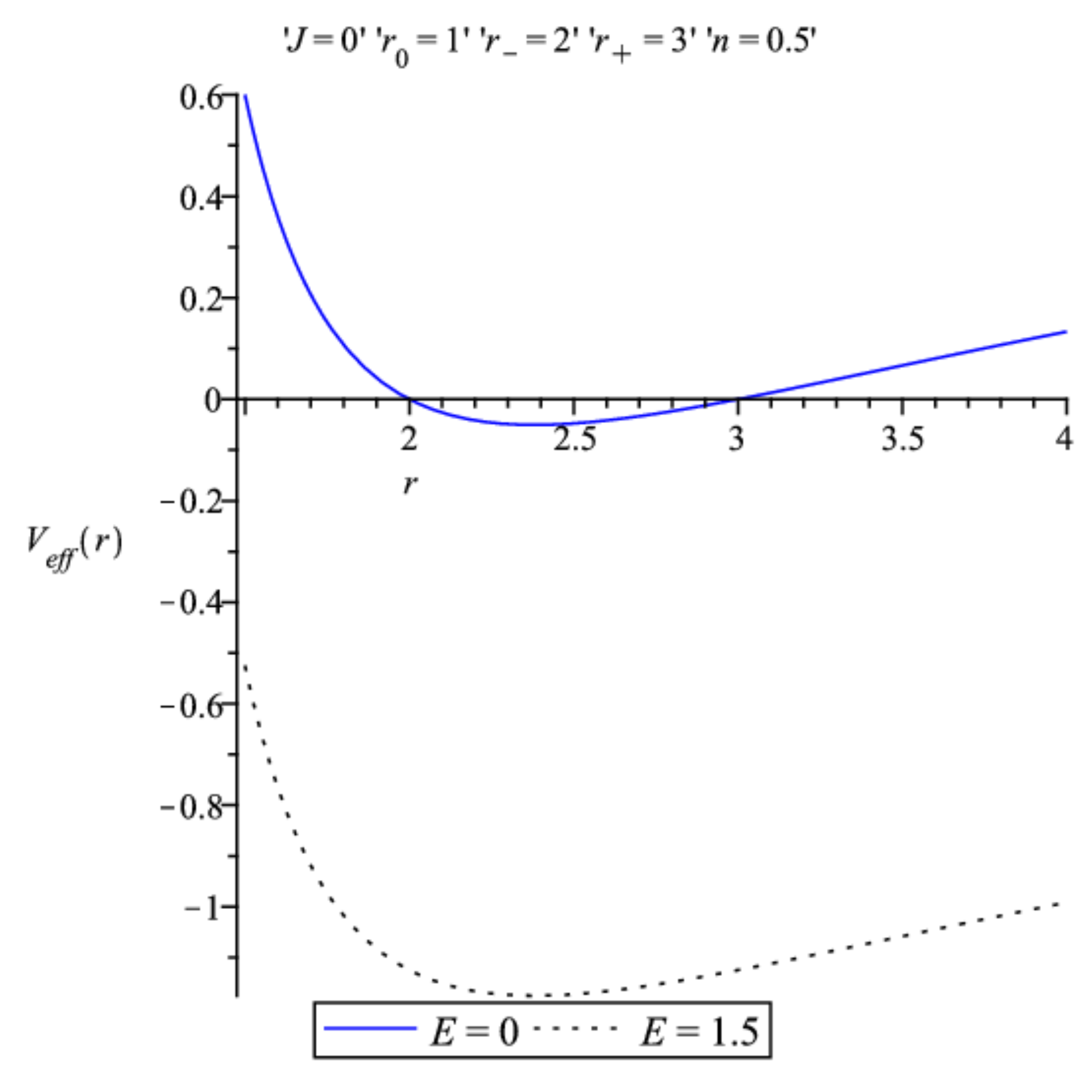}&
\includegraphics[width=7.cm]{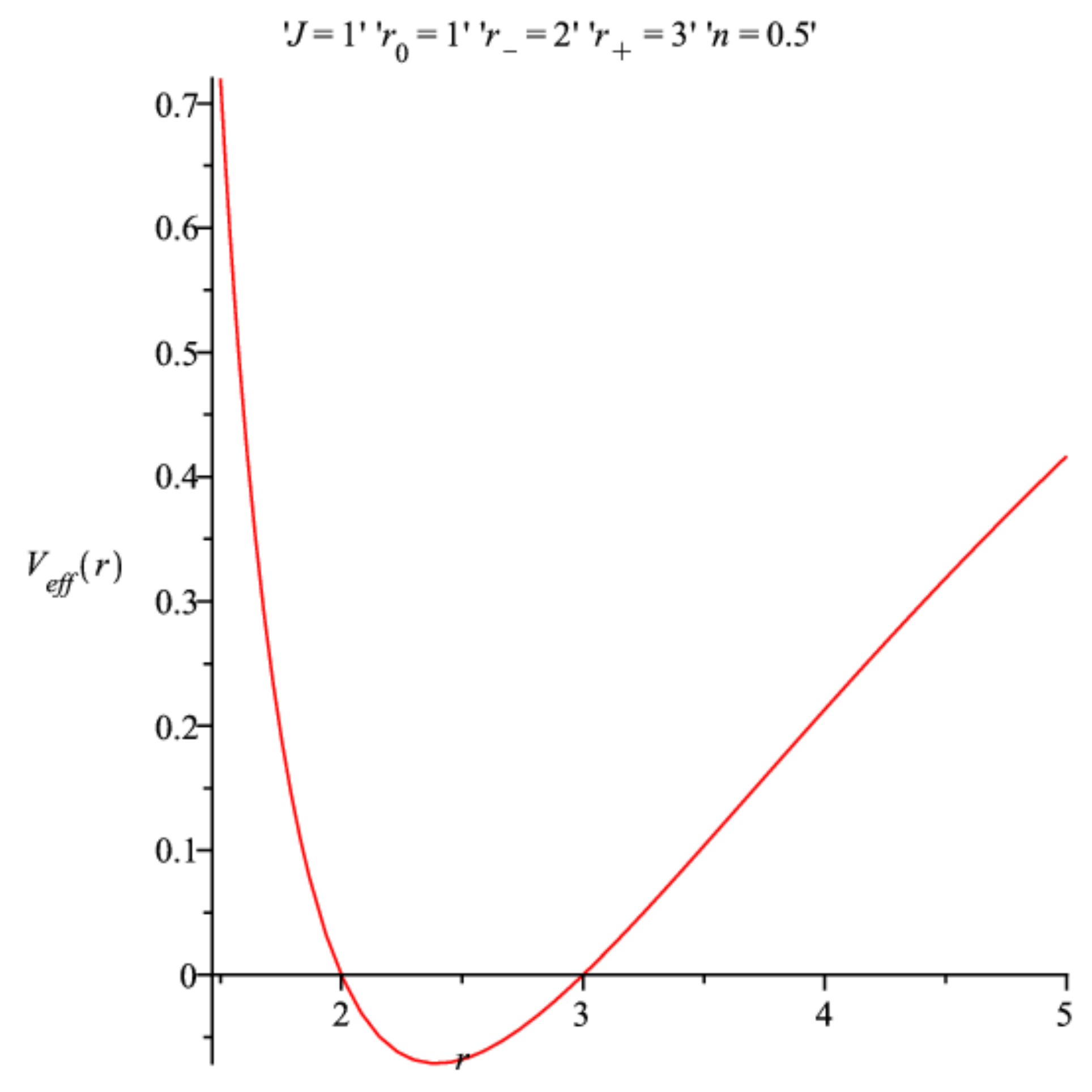}
\\
\end{tabular}
\end{center}
Fig.4: Graphs of $V_{\text{\scriptsize{eff}}}-r$ with $J=0$, $E=0$ and  $E=1.5$ (left) and $V_{\text{\scriptsize{eff}}}-r$ with $J= 1$ and $E=0$ (right), for a massive particle.
\end{figure*}

From Eq. (28) we notice that the solutions for the effective potential coincide with the horizon values for radial geodesics with $E=0$, which is demonstrated graphically in Fig. 4 (left). Further, from Fig. 4 (left) one can see that the motion of the particle is bounded in the interior region of the black hole. The behaviour of the effective potential for $E\neq 0$ is depicted in the bottom part of Fig. 4 (left). In this case, we also deduce that a bound orbit is possible for the massive particle.

Next we will consider the motion of a test particle with non-zero angular momentum. For  $E=0$, the roots of the  effective potential  coincide with the horizons (see Fig. 4 (right)). Thus the particle is bounded in the interior region of the black hole.

\section{Motion of a test Particle}

In this section we study the motion of a test particle of mass $M$ and charge $e$ in the gravitational field of a charged black hole in generalized dilaton-axion gravity. The Hamilton-Jacobi equation [11] is
\begin{equation}
g^{ik}\left(\frac{\partial S}{\partial x^i}+e A_i\right)\left(\frac{\partial S}{\partial x^k}+e A_k\right)+M^2=0,
\end{equation}
where $g_{ik}$, $ A_i$  are the metric potential and the gauge potential respectively and $S$ is Hamilton's standard  characteristic function.
The explicit form of the Hamilton-Jacobi equation for the line element (2) is
\begin{equation}
-\frac{1}{f(r)}\left(\frac{\partial s}{\partial t}+\frac{eQ}{r}\right)^2+f(r)\left(\frac{\partial S}{\partial r}\right)^2+\frac{1}{h(r)}\left(\frac{\partial S}{\partial \theta}\right)^2+\frac{1}{h(r)\sin^2 \theta}\left(\frac{\partial S}{\partial \phi}\right)^2+M^2=0,
\end{equation}
where $Q$ is the charge of the black hole.
 To solve the above partial differential equation, let us assume a separable solution in the form
\begin{equation}
S(t,r,\theta,\phi)=-Et+S_1(r)+S_2(\theta)+J\phi,
\end{equation}
where $E$ and $J$ are the energy and angular momentum of the particle respectively.
After some simplification we obtain
\begin{equation}
S_1(r)=\epsilon \int \left[\frac{\left(E-\frac{eQ}{r}\right)^2}{f^2}-\frac{p^2}{fh}-\frac{M^2}{f}\right]^\frac{1}{2},
\end{equation}
\begin{equation}
S_2(\theta)=\epsilon \int \left(p^2-J^2\text{cosec}^2\theta \right)^\frac{1}{2},
\end{equation}
where $\epsilon=\pm 1$ and $p$ is the separation constant also known as the momentum of the particle.

The radial velocity of the particle is given by
\begin{equation}
\frac{dr}{dt}=f^2\left(E-\frac{eQ}{r}\right)^{-1}\left[\frac{1}{f^2}\left(E-\frac{eQ}{r}\right)^2-\frac{p^2}{fh}-\frac{M^2}{f}\right]^\frac{1}{2}.
\end{equation}
The turning points of the trajectory are obtained by the vanishing of the radial velocity, $\frac{dr}{dt}=0$, which yields
\begin{equation}
\left(E-\frac{eQ}{r}\right)^2-\frac{p^2f}{h}-M^2f=0.
\end{equation}
After solving this equation for $E$, we get
\begin{equation}
E=\frac{eQ}{r}+\sqrt{f}\left(\frac{p^2}{h}+M^2\right)^\frac{1}{2}\label{e}.
\end{equation}
The effective potential is obtained from the relation $V(r)=\frac{E}{M}$ as follows:
\begin{equation}
V=\frac{eQ}{Mr}+\sqrt{f}\left(\frac{p^2}{M^2h}+1\right)^\frac{1}{2}.\label{ep}
\end{equation}
Using Eqs. (3) and (4) the effective potential becomes
\begin{equation}
V(r)=\frac{eQ}{Mr}+\left(1+\frac{p^2(r-r_0)^{2n-2}}{M^2(r+r_0)^{2n}}\right)\left(\frac{\sqrt{(r-r_)(r-r_+)}}{(r-r_0)^{1-n}(r+r_0)^n}\right).
\end{equation}

In the stationary system ($\frac{dV}{dr}=0$) we obtain
\begin{equation*}
-\frac{eQ}{Mr^2}+\frac{1}{2}\frac{\left((r-r_-)(r-r_+)\right)^\frac{1}{2}\left(\frac{p^2(r-r_0)^{2n-2}(2n-2)}{M^2(r-r_0)(r+r_0)^{2n}}-\frac{2p^2n(r-r_0)^{2n-2}}{M^2(r+r_0)^{2n+1}}\right)}{\left(1+\frac{p^2(r-r_0)^{2n-2}}{M^2(r+r_0)^{2n}}\right)^\frac{1}{2}(r-r_0)^{1-n}(r+r_0)^{n}}
\end{equation*}
\begin{equation*}
+\frac{1}{2}\frac{(2r-r_+-r_-)\left(1+\frac{p^2(r-r_0)^{2n-2}}{M^2(r+r_0)^{2n}}\right)^\frac{1}{2}}{\left((r-r_-)(r-r_+)\right)^\frac{1}{2}(r-r_0)^{1-n}(r+r_0)^n}
\end{equation*}
\begin{equation*}
-\frac{(1-n)\left(1+\frac{p^2(r-r_0)^{2n-2}}{M^2(r+r_0)^{2n}}\right)^\frac{1}{2}\left((r-r_-)(r-r_+)\right)^\frac{1}{2}}{(r-r_0)^{2-n}(r+r_0)^n}
\end{equation*}
\begin{equation}
-\frac{n\left(1+\frac{p^2(r-r_0)^{2n-2}}{M^2(r+r_0)^{2n}}\right)^\frac{1}{2}\left((r-r_-)(r-r_+)\right)^\frac{1}{2}}{(r-r_0)^{1-n}(r+r_0)^{n+1}}=0\label{sy}.
\end{equation}
In order to use a more simplified equation and thus be able to visualize it by plotting its graph, one may select some specific value for  $n$. Here we choose   $n=0.5$ (since $0 < n < 1$)  and the simplified form of equation (39) is given by
\begin{equation}
\alpha(r):=\frac{2eQ(r^2-r_0^2)^\frac{1}{2}}{Mr^2}+\frac{2r\left((r-r_-)(r-r_+)\right)^\frac{1}{2}}{(r^2-r_0^2)}\left(\frac{p^2}{M^2(r^2-r_0^2)\beta}
+\beta\right)-\frac{(2r-r_+-r_-)\beta}{\left((r-r_-)(r-r_+)\right)^\frac{1}{2}}=0,
\end{equation}
where $\beta=\left(1+\frac{p^2}{M^2(r^2-r_0^2)}\right)^\frac{1}{2}$.

\subsection{Test particle in static equilibrium}

The momentum $p$ must be zero in the static equilibrium system, thus  from Eq.(40) we get
\begin{eqnarray}
\left(4e^2Q^2-M^2(2r_-r_+ +r_+ ^2-r_- ^2)\right)r^8+\left(-4e^2Q^2(r_-+r_+)+4M^2(r_0^2r_-+r_0^2r_+r_-^2r_++r_-r_+^2)\right)r^7\nonumber\\
+\left(4e^2Q^2(r_-r_+-3r_0^2)-2M^2(6r_0^2r_-r_++r_0^2r_-^2+r_-^2r_+^2+r_0^2r_+^2-r_0^4)\right)r^6\nonumber\\
+\left(12e^2Q^2r_0^2(r_-+r_+)+4M^2r_0^2(r_+r_-^2+r_+^2r_-+r_0^2(r_++r_-))\right)r^5\nonumber\\
+\left(12e^2Q^2r_0^2(r_0^2-r_+r_-)-M^2r_0^4(2r_+r_-+r_+^2+r_-^2)\right)r^4\nonumber\\
-12e^2Q^2r_0^4(r_++r_-)r^3+4e^2Q^2r_0^4(3r_+r_--r_0^2)r^2+4e^2Q^2r_0^6(r_++r_-)r-4e^2Q^2r_0^6r_+r_-=0.
\end{eqnarray}
We notice  that the last term of the above equation is negative. So, this equation has at least one positive real root. Consequently, a bound orbit is possible for the test particle, i.e.  the test particle can be trapped by a  charged black hole in generalized dilaton-axion gravity.   In other words,  a  charged black hole in generalized dilaton-axion gravity exerts an attractive gravitational force on matter.

\subsection{Test particle in non-static equilibrium}

\subsubsection{Test particle without charge ($e=0$)}
In this case Eq.(40) becomes
\begin{eqnarray}
&& M^2(r_-+r_+)r^4-2(r_0^2M^2+r_-r_+M^2+p^2)r^3+3p^2(r_+-r_-)r^2\nonumber\\
&&+2(r^4M^2-r_0^2p^2+r_0^2M^2r_+r_--2p^2r_+r_-)r
+r_0^2(r_+p^2+r_-p^2-M^2r_0^2r_+-M^2r_0^2r_-)=0.
\end{eqnarray}
If
$M^2r_0^2r_++M^2r_0^2r_-> r_+p^2+r_-p^2$, one can see  that the last term of the above equation is negative. Therefore, this equation must have at least one positive real root. Consequently,  a bound orbit for the uncharged  test particle is possible.  If
$M^2r_0^2r_++M^2r_0^2r_-= r_+p^2+r_-p^2$, then Eq. (42) changes to a third degree equation with two changes of sign. By Descarte's rule of sign, this equation must have either two positive roots or no positive roots at all. Thus, a  bound orbit for the uncharged test particle may or may not be possible. For $M^2r_0^2r_++M^2r_0^2r_-< r_+p^2+r_-p^2$, Eq. (42) has  two changes of sign. Here, again a bound orbit for the uncharged test particle may or may not be possible.

\subsubsection{Test particle with charge (e$\neq$0)}

For a charged test particle with $n=0.5$,   the stationary system ($\frac{dV}{dr}=0$) yields the form given in Eq.(40). As this equation is algebraically very complicated, we use the graph of $\alpha(r)$ in order to find out whether there exist any real positive roots. From Fig. 5 one can see that, for different values of the test particle's charge, $\alpha(r)$ given by Eq. (40), does not intersect the $r$-axis. Hence, no real positive roots are possible. As a result, no bound orbit for the  charged test particle is possible.

\begin{figure*}[thbp]
\begin{center}
\includegraphics[width=8 cm]{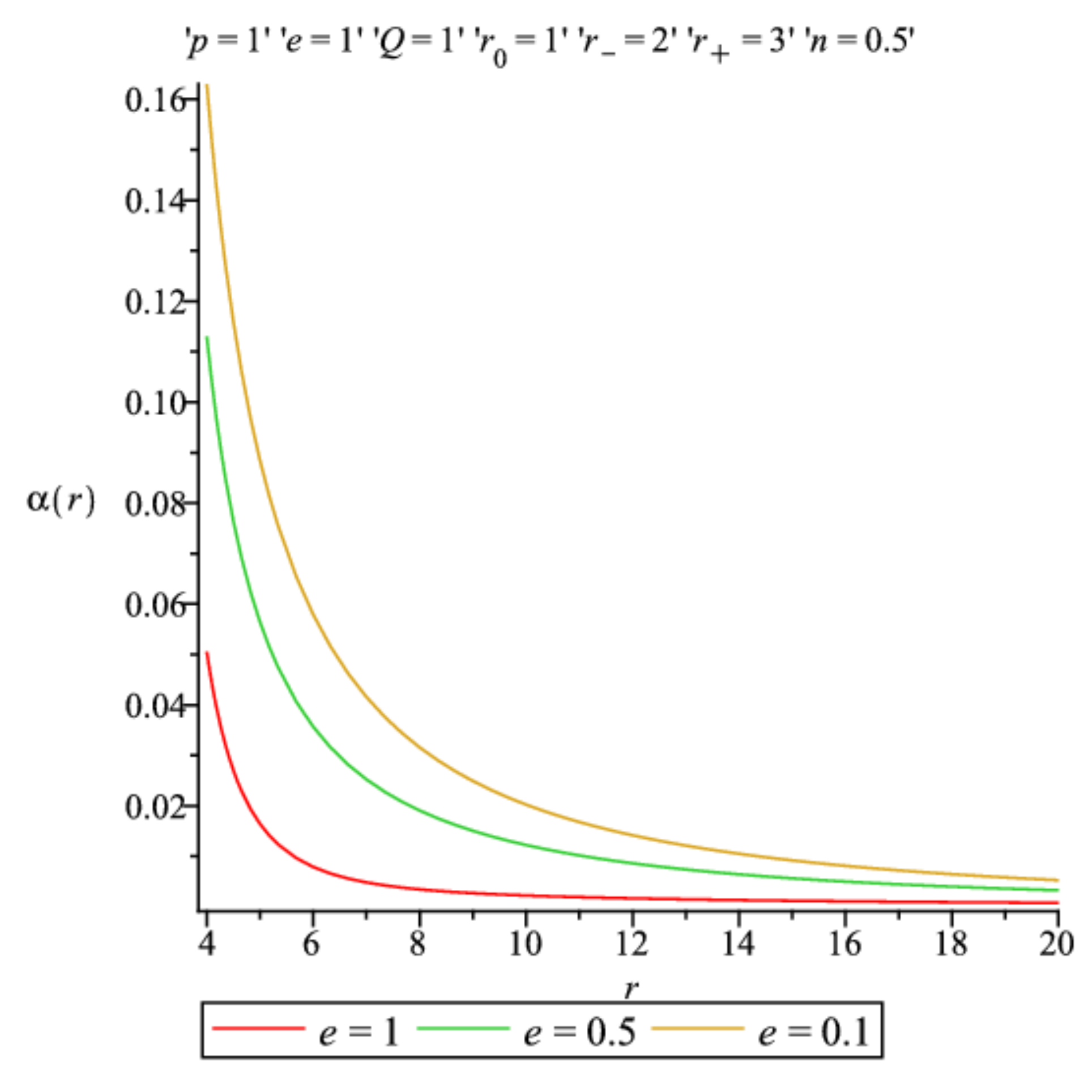}
\end{center}
Fig.5: Graph of $\alpha(r)$ given in Eq. (40) for different values of the test particle's charge.
\end{figure*}

%

%

%

%

%

%

%
\section{Discussion }

 In the present investigation,   we have analyzed the behaviour of  massless and massive particles in the gravitational field of  a charged black hole in generalized dilaton-axion gravity in four dimensions. To this purpose, we have studied the motion of a massless particle ($L=0$) and a massive particle ($L=-1$). We have plotted the graphs of $t$ and the proper time $\tau$ with respect to the radial coordinate $r$. For the massless particle $t$ increases non-linearly with $r$ (Fig. 1 (left)), while the proper time $\tau$ increases linearly with $r$ (Fig. 1 (right)). In the case of the massive particle motion both $t$ and $\tau$ decrease non-linearly with $r$ (Fig. 2).

Further, studying the effective potential we ended up with Eq. (24) from which we conclude that $V_{\text{\scriptsize{eff}}}$ depends on the energy per unit mass $E$ and the angular momentum $J$. For a massless test particle we found that $V_{\text{\scriptsize{eff}}}=-\frac{E^2}{2}$ in the case of radial geodesics ($J=0$), while if $E=0$ the particle behaves like a free particle. In the case of circular geodesics ($J\neq 0$), $V_{\text{\scriptsize{eff}}}$ is given by Eq. (26). In Fig. 3 the behaviour of $V_{\text{\scriptsize{eff}}}$ is presented for zero and non-zero $E$ and various values of $J$. From these graphs it can be inferred that for $E\neq 0$ and $J\neq 0$ the effective potential $V_{\text{\scriptsize{eff}}}$ has a minimum and circular orbits are possible, while for $E\neq 0$ and $J=0$ no stable circular orbits can exist. In the case $E=0$ and $J\neq 0$ the effective potential $V_{\text{\scriptsize{eff}}}$ changes sign two times between the horizons and stable circular orbits must exist.

From the examination of the calculated effective potential, Eq. (27) for a massive particle, we have considered the radial geodesics for the cases $J=0$ and $E=0$, $J=0$ and $E\neq 0$, and  $J\neq 0$ and $E=0$. It is seen (Fig. 4) that in the first case the roots of the effective potential coincide with the horizons' positions and the particle's orbit is bound in the black hole's interior. In the second case, a bound orbit is also possible. Finally, in the third case, i.e. when the particle's angular momentum does not vanish, the roots of the effective potential coincide again with the horizons' positions and the particle's orbit is bound again in the black hole's interior.

As a last step we have examined the motion of a massive and charged test particle in the gravitational field of a charged black hole in generalized dilaton-axion gravity by exploiting the Hamilton-Jacobi equation. The latter is set up for the spacetime geometry considered  and is analytically solved by applying additive separation of variables. As a result, the particle's radial velocity and the effective potential are determined in closed form. Then the case of static equilibrium is examined and it is found that the charged test particle may have a bound orbit,  i.e. it can be trapped by a charged black hole in this context or, stated differently, the charged black hole exerts an attractive gravitational force upon the charged particle in generalized dilaton-axion gravity. In the case of non-static equilibrium, we have distinguished between an uncharged  and a charged test particle. In the former case, conditions have been found for the possibility of existence  of the particle's bound orbit. Finally, when the test particle carries a charge it is seen graphically (Fig. 5) that no bound orbit is possible.

An interesting perspective for future work would be the study of the motion of charged or uncharged test particles and the behaviour of geodesics for rotating black hole solutions or for black hole solutions in more than four spacetime dimensions in the context of generalized dilaton-axion gravity.

\section*{Acknowledgement}
FR would like to thank the authorities of the Inter-University Centre for Astronomy and
Astrophysics, Pune, India for providing research facilities. FR and SS are also grateful to
DST-SERB (Grant No.: EMR/2016/000193) and UGC (Grant No.: 1162/(sc)(CSIR-UGC
NET , DEC 2016)), Govt. of India, for financial support respectively.

\section*{References}

[1] T. Ort\'{\i}n, \textit{Gravity and Strings}, Cambridge University Press, Ch. 16, 2004.

[2] S. Sur, S. Das and S. SenGupta, ``Charged black holes in generalized dilaton-axion gravity'', Journal of High Energy Physics, vol. 10, p. 064, 2005.

[3] T. Ghosh and S. SenGupta, ``Thermodynamics of dilaton-axion black holes", Physical Review D vol. 78,  no. 12, p. 124005, 2008.

[4] 	 A. A. Usmani, Z. Hasan, F. Rahaman, Sk. A. Rakib, S. Ray and Peter K.F. Kuhfittig, ``Thin-shell wormholes from charged black holes in generalized dilaton-axion gravity", General Relativity and Gravitation, vol. 42, no. 12, p. 2901, 2010.

[5]  I. Radinschi, F. Rahaman and A. Ghosh, ``On the energy of charged black holes in generalized dilaton-axion gravity",  International Journal of Theoretical Physics, vol. 49, no. 5, p. 943,  2010.

[6]   Z. M. Yang, X.-L. Li and  Y. Gao, ``Entanglement entropy of charged dilaton-axion black hole and quantum isolated horizon'', European Physical Journal Plus, vol. 131, no. 9, p. 304, 2016.

[7]   T. Ghosh and S. SenGupta, ``Study of superradiant instability of a dilaton-axion black hole under scalar perturbation",  European Physics Letters,  vol. 120, no.5, p. 50003, 2017.

[8] A. S. Bhatia and S. Sur,  ``Dynamical system analysis of dark energy models in scalar coupled metric-torsion theories'',  International Journal of Modern Physics D vol. 26, no.13, p. 1750149, 2017. 

[9] S. Sur and A. S. Bhatia, ``Weakly dynamic dark energy via metric-scalar couplings with torsion'',  Journal of Cosmology and Astroparticle Physics, vol. 2017, no. 07, p. 039, 2017. 

[10] M. Kalam, F. Rahaman and S. Mondal, ``Particle motion around tachyon monopole",  General Relativity and Gravitation, vol 40, no. 9, p. 1849, 2008.

[11] S. Chakraborty and F. Rahaman, ``Motion of test particles around gauge monopoles or near cosmic strings considering semiclassical gravitational effects",  International Journal of Modern Physics D, vol. 9, no. 2, p. 155, 2000.

\end{document}